\documentclass[12pt]{article}
\usepackage{amssymb}
\usepackage{amsmath}
\usepackage{graphicx}

\catcode`@=11
\renewcommand{\section}{\@startsection{section}{1}{0pt}{\medskipamount}
{\medskipamount}{\large\bf}} \numberwithin{equation}{section}
\catcode`@=12

\def\beq{\begin{eqnarray}}    
\def\eeq{\end{eqnarray}}      

\def\={\ =\ }

\begin{document}

\begin{titlepage}
\setcounter{page}{0}

\vskip 2.0cm

\begin{center}

{\LARGE\bf  Jacobi - type identities in algebras and superalgebras}

\vspace{18mm}

{\Large P.M. Lavrov$\,{}^{a,b}$, \
O.V. Radchenko$\,{}^{a}$ \ and \
I.V. Tyutin$\,{}^{c}$
}

\vspace{8mm}

\noindent ${}^{a}${\em
Tomsk State Pedagogical University,\\
Kievskaya St.\ 60, 634061 Tomsk, Russia}

\noindent ${}^{b}${\em
Tomsk State University,\\
Lenin Av.\ 36, 634050 Tomsk, Russia}

\vspace{4mm}

$^c${\em  I.E. Tamm Department of Theoretical Physics,\\
P.N. Lebedev Physical Institute,
\\
119991, Leninsky Prospect, 53, Moscow,Russia }

\vspace{18mm}

\begin{abstract}
\noindent We introduce two remarkable identities written in terms of
single commutators and anticommutators for any three elements of
arbitrary associative algebra. One is a consequence of other
(fundamental identity). From the fundamental identity, we derive a
set of four identities (one of which is the Jacobi identity)
represented in terms of double commutators and anticommutators. We
establish that two of the four identities are independent and show
that if the fundamental identity holds for an algebra, then the
multiplication operation in that algebra is associative. We find a
generalization of the obtained results to the super case and give a
generalization of the fundamental identity in the case of arbitrary
elements. For nondegenerate even  symplectic (super)manifolds, we
discuss analogues of the fundamental identity.
\end{abstract}

\end{center}

\vfill \noindent{\sl Emails:} \ lavrov@tspu.edu.ru,
radchenko@tspu.edu.ru, tyutin@td.lpi.ac.ru\\
\noindent {\sl Keywords:} \ associative algebra, associative
superalgebra,
Jacobi identity, symplectic  supermanifolds \\


\end{titlepage}

\section{Introduction}\label{intro}

\noindent

Algebras endowed with a bracket satisfying the Jacobi identity are
currently used extensively in the formulations of classical and
quantum theories. For example, the formulations of Classical
Mechanics and Classical Field Theory are based on symplectic
manifolds endowed with  the Poisson bracket satisfying the Jacobi
identity \cite{Ar,BSh}.  One of the main ingredients in consistent
formulation of quantum mechanics is the commutator of operators,
which satisfies the Jacobi identity. The main object of General
Relativity is the Riemann tensor, which in particular satisfies the
Bianchi identity following from the Jacobi identity for covariant
derivatives. The quantization of dynamical systems with constraints
in the Hamiltonian formalism \cite{BV1,BF,GT,HT} assumes the use of
even symplectic supermanifolds \cite{Leites, DeWitt} endowed with
superextension of the commutator and a Poisson bracket satisfying
 the generalized Jacobi identity. The quantum theory  of gauge fields in the Lagrangian
 formalism is constructed using odd symplectic supermanifolds endowed
 with an
 antibracket (a super Poisson bracket \cite{BV}) that satisfies
the generalized
 Jacobi identity. The list of such descriptions can be continued.

Here, we consider the Jacobi identities naturally existing for any
associative algebra from a new standpoint (in our understanding).
For this, we use  a remarkable identity for any three elements of a
given associative algebra presented in terms of only single
commutators. The Jacobi identity written, as is known, in terms of
double commutators and anticommutators follows from this identity.
Using the anticommutator, we introduce a second (fundamental)
identity for an arbitrary associative algebra written for three
elements of the algebra in terms of single commutators and
anticommutators. We show that the first identity is a consequence of
the second. This allows speaking of this identity as a fundamental
(basic) identity for any associative algebra. For identities (one of
which is the Jacobi identity) in terms  of double commutators and
anticommutators can be derived from the fundamental identity. Among
these identities, two are independent. For any algebra, we prove
that if the fundamental identity is satisfied, then the
multiplication operation is associative. We generalize the basic
relations and statements
 to the case of superalgebras.

This paper is organized as follows. In Section~2, we introduce the
fundamental identity written in terms of single commutators and
anticommutators for arbitrary associative algebras and derive a set
of four (reducible) identities in terms of double commutators and
anticommutators. We show that an algebra endowed with the
fundamental identity is an associative algebra. In Section~3, we
extend the results obtained in the preceding sections to the super
case. In Section~4, we study a new interesting identity that exists
for nondegenerate symplectic (super)manifolds. In Section~5, we
generalize the basic identities discussed in Section 2 to the case
of an arbitrary number of elements.
Finally, we make some concluding concluding remarks in Section~6.\\

\section{Remarkable identities in associative algebras}

\noindent

We consider an arbitrary associative algebra $\cal A$ with elements
$X\in {\cal A}$.  Let $T_i, i=1,2,...,n$ be a basis in $\cal A$.
Then there exists the decomposition $X=x^iT_i$ for any element $X$
in $\cal A$. Because $T_iT_j\in {\cal A} $, we have
\begin{eqnarray}
\label{TT} T_iT_j=F^{\;\;k}_{ij}\;T_k,
\end{eqnarray}
where $F^{\;\;k}_{ij}$ are structure constants of the algebra. In
terms of $F^{\;\;k}_{ij}$, the associativity conditions
$(XY)Z=X(YZ)$ is
\begin{eqnarray}
\label{FFas}
F^{\;\;n}_{ij}F^{\;\;m}_{nk}=F^{\;\;n}_{j\;k}F^{\;\;m}_{in}.
\end{eqnarray}
These constants can be uniquely represented as sum of symmetric and
antisymmetric terms
\begin{eqnarray}
\label{F}
F^{\;\;k}_{ij}=\frac{1}{2}c^{\;\;k}_{ij}+\frac{1}{2}f^{\;\;k}_{ij},
\end{eqnarray}
where  $c^{\;\;k}_{ij}$ and $f^{\;\;k}_{ij}$  have the symmetries
\begin{eqnarray}
\label{fsym} c^{\;\;k}_{ij}= c^{\;\;k}_{j\;i},\quad
f^{\;\;k}_{ij}=-f^{\;\;k}_{j\;i}.
\end{eqnarray}
The commutator $[\cdot,\cdot]$ and anticommutator $\{\cdot,\cdot\}$
in this algebra are defined for any two elements $X,Y\in {\cal A}$
by the relations
\begin{eqnarray}
\label{comm} [X,Y]=XY-YX,\qquad \{X,Y\}=XY+YX,
\end{eqnarray}
which are elements of ${\cal A}$.

The Leibniz rules for the commutator and anticommutator follow from
(\ref{comm}),
\begin{eqnarray}
\label{Lcomm} [X,YZ]=[X,Y]Z+Y[X,Z],\qquad \{X,YZ\}=\{X,Y\}Z-Y[X,Z].
\end{eqnarray}

 It is clear that
\begin{eqnarray}
\label{commTT} [T_i,T_j]=f^{\;\;k}_{ij}T_k ,\qquad \{T_i,T_j\}=
c^{\;\;k}_{ij}T_k.
\end{eqnarray}

The following remarkable identities written in terms of single
commutators and anticommutators exist for any associative algebra
$\cal A$ and for any $X,Y,Z\in {\cal A}$ (without any reference to a
basis):
\begin{eqnarray}
\label{FunI}
&&[X,YZ]+ [Z,XY]+[Y,ZX]\equiv 0,\\
\label{FunII} &&[X,YZ]+\{Y,ZX\}-\{Z,XY\}\equiv 0.
\end{eqnarray}
From these identities, we can derive a set of identities in terms of
double commutators and anticommutators. In particular, the Jacobi
identity
\begin{eqnarray}
\label{JI} [X,[Y,Z]]+ [Z,[X,Y]]+ [Y,[Z,X]]\equiv 0
\end{eqnarray}
follows from (\ref{FunI}). Moreover, we can deduce the identity
containing the anticommutator from (\ref{FunI}),
\begin{eqnarray}
\label{JIab} [X,\{Y,Z\}]+ [Z,\{X,Y\}]+ [Y,\{Z,X\}]\equiv 0
\end{eqnarray}
Similarly, we can derive the following identities from
(\ref{FunII}):

\begin{eqnarray}
\label{JIantiI} &&[X,\{Y,Z\}]- \{Z,[X,Y]\}+ \{Y,[Z,X]\}\equiv 0,\\
  &&[X,[Y,Z]]+\{Y,\{Z,X\}\}-\{Z,\{X,Y\}\}\equiv 0.
\label{JIantiII}
\end{eqnarray}
We note that identities (\ref{FunI}) and (\ref{FunII}) are not
independent, because summing (\ref{FunII}) over cyclic permutations
gives identity (\ref{FunI}). Therefore, we have  reason to regard
identity (\ref{FunII}) as the fundamental identity in associative
algebras because the identities (\ref{FunI}), (\ref{JI}),
(\ref{JIab}), (\ref{JIantiI}) and (\ref{JIantiII}) can be derived
from it. In turn, the set of identities (\ref{JI})-(\ref{JIantiII})
is not independent. Indeed, summing of (\ref{JIantiI}) over cyclic
permutations gives identity (\ref{JIab}). The same operation applied
to identity (\ref{JIantiII}) leads to  Jacobi identity (\ref{JI})
\cite{MR}. We note that identities (\ref{JIantiI}), (\ref{JIantiII})
were also discussed  for associative algebras in \cite{LV}. It is
clear that having identities (\ref{JIantiI})-- (\ref{JIantiII}) and
explicit realization of commutator and anticommutator (\ref{comm}),
we can reproduce fundamental identity (\ref{FunII}).

We note  that  from  associativity condition (\ref{FFas}), we can
derive analogues of identities (\ref{FunI})-(\ref{JIantiII}) in
terms of structure constants. In particular, the Jacobi identities
for the antisymmetric parts $f^{\;\;k}_{ij}$ of the structure
constants have the form
\begin{eqnarray}
\label{JIf} f^{\;\;m}_{ij}f^{\;\;n}_{mk}+
f^{\;\;m}_{ki}f^{\;\;n}_{mj}+ f^{\;\;m}_{jk}f^{\;\;n}_{mi} \equiv 0.
\end{eqnarray}
The identities containing symmetric and antisymmetric parts of
structure constants can be written as
\begin{eqnarray}
\nonumber \label{JIfc}
&&c^{\;\;m}_{ij}f^{\;\;n}_{mk}+
c^{\;\;m}_{ki}f^{\;\;n}_{mj}+ c^{\;\;m}_{jk}f^{\;\;n}_{mi} \equiv 0,\\
\label{JIfcII}&&f^{\;\;m}_{ij}c^{\;\;n}_{mk}-f^{\;\;m}_{ki}c^{\;\;n}_{mj}+
c^{\;\;m}_{jk}f^{\;\;n}_{mi}\equiv 0,\\
\nonumber
 \label{JIfcIII}&& c^{\;\;m}_{ij}c^{\;\;n}_{mk}-
c^{\;\;m}_{ki}c^{\;\;n}_{mj} +f^{\;\;m}_{jk}f^{\;\;n}_{mi}\equiv 0.
\end{eqnarray}

The existence of fundamental identity (\ref{FunII}) allows
discussing the associativity of multiplication in the algebra from a
new standpoint. Indeed, let ${\cal A}$ be an algebra, then we can
introduce the commutator and anticommutator by  rules (\ref{comm})
as elements of ${\cal A}$. We suppose that fundamental identity
(\ref{FunII}) is satisfied. We know that identities
(\ref{JIab})-(\ref{JIantiII}) follow from (\ref{FunII}). We can then
show that the multiplication in ${\cal A}$ is associative. For this,
we introduce the multiplication  for any two elements $X,Y\in {\cal
A}$ as
\begin{eqnarray}
\label{defmula} XY=\frac{1}{2}\;\Big([X,Y]+\{X,Y\}\Big)
\end{eqnarray}
and we verify that equality
\begin{eqnarray}
\label{Imul} (XY)Z=X(YZ)
\end{eqnarray}
holds. Indeed, from (\ref{defmula}) that
\begin{eqnarray}
&&(XY)Z=[[X,Y],Z]+[\{X,Y\},Z]+\{[X,Y],Z\}+\{\{X,Y\},Z\}, \label{ass1}\\
&&X(YZ)=[X,[Y,Z]]+[X,\{Y,Z\}]+\{X,[Y,Z]\}+\{X,\{Y,Z\}\}.
\label{ass2}
\end{eqnarray}
For the difference between (\ref{ass1}) and (\ref{ass2}), we obtain
\begin{eqnarray}\label{ass3}
&&(XY)Z-X(YZ)=-[Z,[X,Y]]-[X,[Y,Z]]- \nonumber \\
&&-[Z,\{X,Y\}]-[X,\{Y,Z\}]+\{Z,[X,Y]\}-\{X,[Y,Z]\}+ \nonumber \\
&&+\{Z,\{X,Y\}\}-\{X,\{Y,Z\}\}.\label{ass3}
\end{eqnarray}
With identity (\ref{JIantiII}) and Jacobi identity (\ref{JI}), which
follows from it, taking into account, relation (\ref{ass3}) becomes
\begin{equation}\label{ass4}
(XY)Z-X(YZ)=-[Z,\{X,Y\}]-[X,\{Y,Z\}]+\{Z,[X,Y]\}-\{X,[Y,Z]\}.
\end{equation}
Expressing $[Z,\{X,Y\}]$ and $[X,\{Y,Z\}]$ using identity
(\ref{JIantiI}), we now obtain associativity condition (\ref{Imul}).
Therefore, for any algebra ${\cal A}$, fundamental identity
(\ref{FunII}) is equivalent to the associativity condition. We
formulate this result as a theorem:

{\bf Theorem 1.} {\it For the associativity of multiplication in an
algebra ${\cal A}$,  it is necessary and sufficient that  identity
(\ref{FunII}), where the commutators and anticommutators are defined
by rule (\ref{comm}), is satisfied}.

Any associative algebra has the natural structure of a Lie algebra
if the Lie bracket is defined in terms of associative
multiplication.  The converse does not hold in general: the Lie
bracket in the general case does not allow  introducing an
associative multiplication. But the following  theorem holds.

{\bf Theorem 2.} {\it For any algebra ${\cal A}$ equipped with two
bilinear operations $[\cdot,\cdot]$ and $\{\cdot,\cdot\}$ with the
symmetry properties}:
\begin{eqnarray}
\label{syma} [X,Y]=-[Y,X], \qquad \{X,Y\}=\{Y,X\},
\end{eqnarray}
{\it and satisfying the identities (\ref{JIantiI}) and
(\ref{JIantiII}), associative multiplication can be introduced by
rule (\ref{defmula})}.

The proof of  Theorem 2 is in fact contained  in relations
(\ref{Imul}) - (\ref{ass4}).

We note that the above results do not use assumptions about the
finiteness or infiniteness of a system of basis vectors $\{T_i\}$ of
the algebra $\cal A$ or even the existence of a basis at all.
\\

\section{Superextension}

\noindent

The results obtained in the preceding section can be extended to any
associative superalgebras ${\cal A}_s$ with elements $X\in {\cal
A}_s$ having the Grassmann parity $\varepsilon(X)$. Let $T_i,
i=1,2,...,n,\; \varepsilon(T_i)=\varepsilon_i$ be a basis in ${\cal
A}_s$ such that there exist the decomposition $X=x^iT_i,\;
\varepsilon(x^i)=\varepsilon(X)+\varepsilon_i$, for any element
$X\in {\cal A}_s$. Because $T_iT_j\in {\cal A}_s $, we have
\begin{eqnarray}
\label{TTs} T_iT_j=F^{\;\;k}_{ij}\;T_k,
\end{eqnarray}
where $F^{\;\;k}_{ij},\;(\varepsilon(F^{\;\;k}_{ij})=
\varepsilon_i+\varepsilon_j+\varepsilon_k)$, are structure
constants. In terms of $F^{\;\;k}_{ij}$ the associativity condition
$(XY)Z=X(YZ)$ is
\begin{eqnarray}
\label{FFassup}
F^{\;\;n}_{ij}F^{\;\;m}_{nk}=F^{\;\;n}_{j\;k}F^{\;\;m}_{in}
(-1)^{\varepsilon_i(\varepsilon_j+\varepsilon_k+\varepsilon_n)}.
\end{eqnarray}

In the super case, the commutator $[\cdot,\cdot]$ and anticommutator
$\{\cdot,\cdot\}$ are introduced for any two elements $X,Y\in {\cal
A}_s$ by the relations
\begin{eqnarray}
\label{scomm} [X,Y]=XY-(-1)^{\varepsilon(X)\varepsilon(Y)}YX,\qquad
\{X,Y\}=XY+(-1)^{\varepsilon(X)\varepsilon(Y)}YX
\end{eqnarray}
with obvious  symmetries properties
\begin{eqnarray}
\label{scommsym}
[X,Y]=-[Y,X](-1)^{\varepsilon(X)\varepsilon(Y)},\qquad
\{X,Y\}=\{Y,X\}(-1)^{\varepsilon(X)\varepsilon(Y)}.
\end{eqnarray}
 The Leibniz rules follow from (\ref{scomm})
\begin{eqnarray}
\nonumber \label{Lscomm}
&&[X,YZ]=[X,Y]Z+Y[X,Z](-1)^{\varepsilon(X)\varepsilon(Y)},\\
\label{Lsanticomm}
&&\{X,YZ\}=\{X,Y\}Z-Y[X,Z](-1)^{\varepsilon(X)\varepsilon(Y)}.
\end{eqnarray}
For basis elements, we have
\begin{eqnarray}
\label{scommTT} [T_i,T_j]=f^{\;\;k}_{ij}T_k ,\qquad \{T_i,T_j\}=
c^{\;\;k}_{ij}T_k,
\end{eqnarray}
where structure constants $f^{\;\;k}_{ij}$ and $c^{\;\;k}_{ij}$ have
the symmetry properties
\begin{eqnarray}
\label{sfsym} c^{\;\;k}_{ij}=
c^{\;\;k}_{ji}(-1)^{\varepsilon_i\varepsilon_j},\quad
f^{\;\;k}_{ij}=-f^{\;\;k}_{ji}(-1)^{\varepsilon_i\varepsilon_j}.
\end{eqnarray}
They can be identified with symmetric and antisymmetric parts of the
structure coefficients $F^{\;\;k}_{ij}$
\begin{eqnarray}
\label{Fcfind} c^{\;\;k}_{ij}=F^{\;\;k}_{ij}+F^{\;\;k}_{j\;i}
(-1)^{\varepsilon_i\varepsilon_j},\quad
f^{\;\;k}_{ij}=F^{\;\;k}_{ij}-F^{\;\;k}_{j\;i}
(-1)^{\varepsilon_i\varepsilon_j}.
\end{eqnarray}

For any associative superalgebra  ${\cal A}_s$ and for any $X,Y,Z\in
{\cal A}_s$, we have the identities
\begin{eqnarray}
\label{FunIs} &&[X,YZ](-1)^{\varepsilon(X)\varepsilon(Z)}+
[Z,XY](-1)^{\varepsilon(Z)\varepsilon(Y)}+
[Y,ZX](-1)^{\varepsilon(Y)\varepsilon(X)}\equiv 0\\
\label{FunIIs} &&[X,YZ](-1)^{\varepsilon(X)\varepsilon(Z)}+
\{Y,ZX\}(-1)^{\varepsilon(Y)\varepsilon(X)}-
\{Z,XY\}(-1)^{\varepsilon(Z)\varepsilon(Y)}\equiv 0 ,
\end{eqnarray}
which generalize relations (\ref{FunI}) and (\ref{FunII}). Identity
(\ref{FunIs}) can be derived from (\ref{FunIIs}) and can be regarded
 as the fundamental identity for
associative superalgebras. A set of identities in terms of double
commutators and anticommutators is
\begin{eqnarray}
\label{JIs} [X,[Y,Z]](-1)^{\varepsilon(X)\varepsilon(Z)}+
[Z,[X,Y]](-1)^{\varepsilon(Z)\varepsilon(Y)}+
[Y,[Z,X]](-1)^{\varepsilon(Y)\varepsilon(X)}\equiv 0,
\end{eqnarray}
\begin{eqnarray}
\label{JIsI} [X,\{Y,Z\}](-1)^{\varepsilon(X)\varepsilon(Z)}+
[Z,\{X,Y\}](-1)^{\varepsilon(Z)\varepsilon(Y)}+
[Y,\{Z,X\}](-1)^{\varepsilon(Y)\varepsilon(X)}\equiv 0,
\end{eqnarray}
\begin{eqnarray}
\nonumber \label{JIantis}
&&[X,\{Y,Z\}](-1)^{\varepsilon(X)\varepsilon(Z)}-
\{Z,[X,Y]\}(-1)^{\varepsilon(Z)\varepsilon(Y)}+\\
&&\qquad\qquad\qquad\qquad\qquad +
\{Y,[Z,X]\}(-1)^{\varepsilon(Y)\varepsilon(X)}\equiv 0,\\
\nonumber \label{JIantiIs}
&&[X,[Y,Z]](-1)^{\varepsilon(X)\varepsilon(Z)}+
\{Y,\{Z,X\}\}(-1)^{\varepsilon(X)\varepsilon(Y)}-\\
 &&\qquad\qquad \qquad\qquad\qquad-
\{Z,\{X,Y\}\}(-1)^{\varepsilon(Z)\varepsilon(Y)} \equiv 0.
\end{eqnarray}
Identities (\ref{JIs}) and (\ref{JIsI}) respectively follow from
(\ref{JIantiIs}) and (\ref{JIantis}) by summing over cyclic
permutations.

The identities in terms of symmetric and antisymmetric parts of
structure constants follow from associativity condition
(\ref{FFassup}),
\begin{eqnarray}
\nonumber \label{JIfs}
&&f^{\;\;n}_{ij}f^{\;\;m}_{nk}(-1)^{\varepsilon_i\varepsilon_k}+
f^{\;\;n}_{ki}f^{\;\;m}_{nj}(-1)^{\varepsilon_k\varepsilon_j}+
f^{\;\;n}_{j\;k}f^{\;\;m}_{ni}(-1)^{\varepsilon_j\varepsilon_i}
\equiv
0,\\
\nonumber \label{JIfcs}
&&c^{\;\;n}_{ij}f^{\;\;m}_{nk}(-1)^{\varepsilon_i\varepsilon_k}+
c^{\;\;n}_{ki}f^{\;\;m}_{nj}(-1)^{\varepsilon_j\varepsilon_k}+
 c^{\;\;n}_{j\;k}f^{\;\;m}_{ni}(-1)^{\varepsilon_i\varepsilon_j} \equiv 0,\\
\label{JIfcIIs}&&f^{\;\;n}_{ij}c^{\;\;m}_{nk}(-1)^{\varepsilon_i\varepsilon_k}
-f^{\;\;n}_{ki}c^{\;\;m}_{nj}(-1)^{\varepsilon_k\varepsilon_j}+
c^{\;\;n}_{j\;k}f^{\;\;m}_{ni}(-1)^{\varepsilon_i\varepsilon_j}\equiv 0,\\
\nonumber \label{JIfcIIIs}&&
c^{\;\;n}_{ij}c^{\;\;m}_{nk}(-1)^{\varepsilon_i\varepsilon_k}-
c^{\;\;n}_{ki}c^{\;\;m}_{nj}(-1)^{\varepsilon_j\varepsilon_k}
+f^{\;\;n}_{j\;k}f^{\;\;m}_{ni}(-1)^{\varepsilon_i\varepsilon_j}\equiv
0.
\end{eqnarray}

We can again consider the associativity of multiplication operation
in superalgebras from a new standpoint. For this, we consider a
superalgebra ${\cal A}_s$. We introduce the commutator and
anticommutator by rule (\ref{scomm}) as elements of ${\cal A}_s$. We
suppose that fundamental identity (\ref{FunIIs}) is satisfied. Using
the representation of the  multiplication for any two elements
$X,Y\in {\cal A}_s$ in the form
\begin{eqnarray}\label{defmuls}
XY=\frac{1}{2}\Big([X,Y]+\{X,Y\}\Big)
\end{eqnarray}
and repeating the proof given in Section 2, we obtain the
associativity of multiplication
\begin{eqnarray}
(XY)Z=X(YZ).
\end{eqnarray}
Consequently, we have the following theorem.

{\bf Theorem 3.} {\it For associativity of multiplication in a
superalgebra ${\cal A}_s$, it is necessary and sufficient that
identity (\ref{FunIIs}),  where the commutators and anticommutators
are defined by rule (\ref{scomm}), is satisfied}.

Any associative superalgebra has a natural structure of a Lie
superalgebra if the Lie superbracket is defined in terms of
associative multiplication by the formula
$[X,Y]=XY-YX(-1)^{\varepsilon(X)\varepsilon(Y)}$. The converse
 does not hold, generally speaking: the Lie superbracket in the
 general case does not allow introducing an associative multiplication. But the following theorem
 holds.

{\bf Theorem 4.} {\it Let ${\cal A}_s$ be  a superalgebra equipped
with two bilinear operations $[\cdot,\cdot]$ and $\{\cdot,\cdot\}$
with  the symmetry properties}:
\begin{eqnarray}
[X,Y]=-(-1)^{\varepsilon(X)\varepsilon(Y)}[Y,X],\qquad
\{X,Y\}=(-1)^{\varepsilon(X)\varepsilon(Y)}\{Y,X\}\;.
\end{eqnarray}
{\it If these operations satisfy  identities (\ref{JIantis}) and
(\ref{JIantiIs}), then  an associative multiplication can be
introduce in ${\cal A}_s$.}

Indeed, we define the multiplication $X\circ Y$ by  rule
(\ref{defmuls}), \[ X\circ Y=\frac{1}{2}\Big([X,Y]+\{X,Y\}\Big),
\] and apply the proof given in Section 2. We then obtain
the associativity of this multiplication. In terms of this
multiplication, the binary operations introduced above have the
usual representation
\begin{eqnarray}
[X,Y]=X\circ Y -Y\circ X(-1)^{\varepsilon(X)\varepsilon(Y)},\quad
\{X,Y\} =X\circ Y +Y\circ X(-1)^{\varepsilon(X)\varepsilon(Y)}.
\end{eqnarray}
We note that the above consideration does not contain  any
assumptions concerning a basis in the superalgebra.
\\

\section{Nondegenerate symplectic (super)manifolds}

We consider a nondegenerate symplectic supermanifold $({\cal M},
\omega)$, where ${\cal M}$ is a supermanifold and $\omega$ is a
nondegenerate closed 2-form with the Grassmann parity
$\varepsilon(\omega(X,Y))=\varepsilon(X)+\varepsilon(Y)+\varepsilon(\omega)$,
where $X$ and $Y$ are elements of the cotangent space of ${\cal M}$.
We speak of an even symplectic supermanifold if
$\varepsilon(\omega)=0$ and of an odd symplectic supermanifold if
$\varepsilon(\omega)=1$. It is well known (see, e. g., \cite{Ar})
that an even symplectic supermanifold is the foundation for
describing classical dynamical systems in the Hamiltonian formalism.
In turn, the covariant quantization of gauge theories
(Batalin-Vilkovisky method \cite{BV}) is based on a nondegenerate
odd symplectic supermanifolds. Any nondegenerate closed symplectic
structure defines the  Poisson superbracket $\{F,G\}$\;
$(\varepsilon(\{F,G\})=\varepsilon(F)+\varepsilon(G)+\varepsilon(\omega))$,
which for any two scalar functions $F,\;G$ on ${\cal M}$ is a scalar
under general changes of coordinates on ${\cal M}$. The  Poisson
superbracket has the properties of antisymmetry
\begin{eqnarray}
\label{PBant} \{F,G\}=-\{G,F\}
(-1)^{(\varepsilon(F)+\varepsilon(\omega))\;(\varepsilon(G)+\varepsilon(\omega))}
\end{eqnarray}
the linearity
\begin{eqnarray}
\{F+G,H\}=\{F,H\}+\{G,H\}
\end{eqnarray}
and satisfied the Leibniz rule
\begin{eqnarray}
\label{PBL}
\{F,GH\}=\{F,G\}H+\{F,H\}G(-1)^{\varepsilon(H)\varepsilon(G)}
\end{eqnarray}
and the Jacobi identity
\begin{eqnarray}
\nonumber
&&\{F,\{G,H\}\}(-1)^{(\varepsilon(F)+\varepsilon(\omega))\;(\varepsilon(H)+\varepsilon(\omega))}+
\{H,\{F,G\}\}(-1)^{(\varepsilon(H)+\varepsilon(\omega))\;(\varepsilon(G)+\varepsilon(\omega))}+\\
\label{PBJI}
&&+\{G,\{H,F\}\}(-1)^{(\varepsilon(G)+\varepsilon(\omega))\;
(\varepsilon(F)+\varepsilon(\omega))}\equiv 0,
\end{eqnarray}
which is consequence of the closedness of symplectic structure.

In even case ($\varepsilon(\omega)=0$), the  Poisson superbracket
coincides with the superextension of the Poisson bracket. In odd
case ($\varepsilon(\omega)=1$), the Poisson superbracket is the
antibracket, which is one of the fundamental operations in the BV
quantization method \cite{BV,BV5} and it is known in mathematics as
the Buttin bracket \cite{Bu}.

 We note that in even case ($\varepsilon(\omega)=0$), the identity
\begin{eqnarray}
\label{PBnI} \{F,GH\}(-1)^{\varepsilon(F)\varepsilon(H)}+
\{H,FG\}(-1)^{\varepsilon(H)\varepsilon(G)}+
\{G,HF\}(-1)^{\varepsilon(G)\varepsilon(F)} \equiv 0.
\end{eqnarray}
follows from (\ref{PBant})
 and (\ref{PBL}).
Unfortunately, in contrast to associative algebras, we cannot regard
this identity as  fundamental for nondegenerate closed even
symplectic supermanifolds, because we cannot deduce Jacobi identity
(\ref{PBJI}) from (\ref{PBnI}). Nevertheless if the canonical
quantization is applied to a dynamical system for which the phase
space is described by a nondegenerate even closed symplectic
supermanifold, then the Poisson bracket is transformed into the
commutator $\{F,G\}\rightarrow (i\hbar)^{-1}[{\hat F},{\hat G}]$,
and the identity (\ref{PBnI}) reduces to (\ref{FunIs}) for the
operators ${\hat F}, {\hat G}, {\hat H}$.

\section{Generalization of basic identities}

We note that for any associative algebra, we have the identities
\begin{eqnarray}
\label{GenId} [X_1,X_2X_3\cdots X_n]+cycle(X_1,X_2,...,X_n) \equiv
0,\qquad n=3,4,...,
\end{eqnarray}
which generalize identity (\ref{FunI}). The Jacobi identity
\begin{eqnarray}
\nonumber [[X_1,X_2],X_3]+ [[X_3,X_1],X_2]+ [[X_2,X_3],X_1]\equiv 0
\end{eqnarray}
follows from (\ref{GenId}) for $n=3$
\begin{eqnarray}
\label{GenJI3our} &&[X_1,X_2X_3]- [X_2,X_1X_3]+cycle(X_1,X_2,X_3)
\equiv 0.
\end{eqnarray}

The generalized Jacobi identity for $n=4$ was discussed in \cite{We}
and had the form
\begin{eqnarray}
\label{GenJI4} [[[X_1,X_2],X_3],X_4]+ [[[X_2,X_1],X_4],X_3]+
[[[X_3,X_4],X_1],X_2]+[[[X_4,X_3],X_2],X_1] \equiv 0.
\end{eqnarray}
This  identity can be obtained from (\ref{GenId}). Indeed, a direct
verification shows that it in fact coincides with
\begin{eqnarray}
\label{GenJI4our1} &&[X_1,X_2X_3X_4]- [X_2,X_1X_3X_4]+
[X_4,X_3X_2X_1]-
[X_4,X_3X_1X_2]+\\
\nonumber &&\qquad+ cycle(X_1,X_2,X_3,X_4) \equiv 0.
\end{eqnarray}
The generalization of identity (\ref{GenJI4}) for $n=5,6,...$ was
given in \cite{BL}, and  these generalized Jacobi identities can
again be derived from fundamental identities (\ref{GenId}).

We can also suggest a generalization of   identity (\ref{FunII}).
For four elements, we have
\begin{eqnarray}
\label{Funm4}
[X_1,X_2X_3X_4]-\{X_4,X_1X_2X_3\}+\{X_3,X_4X_1X_2\}+[X_2,X_3X_4X_1]\equiv
0.
\end{eqnarray}

In general, for any associative algebra, there exist the identities
\begin{eqnarray}
\label{FunmII}
&&[X_1, X_2X_3\cdots X_n]-\{X_n,X_1X_2\cdots X_{n-1}\}+\{X_{n-1},X_nX_1\cdots X_{n-2}\}+\\
\nonumber
&&+[X_{n-2},X_{n-1}X_nX_1\cdots X_{n-3}]+[X_{n-3}, X_{n-2}X_{n-1}X_nX_1\cdots X_{n-4}]+\\
\nonumber &&+\cdots +[X_2,X_3\cdots X_nX_1]\equiv 0,\quad n\geq 4 .
\end{eqnarray}
The proof of (\ref{FunmII}) is based on the obvious identity
\begin{eqnarray}
\label{FunmuII}
&&[X_n,X_1 X_2\cdots X_{n-1}]+[X_{n-1},X_nX_1\cdots X_{n-2}]+\\
\nonumber &&+\{X_n,X_1X_2\cdots X_{n-1}\}-\{X_{n-1},X_nX_1\cdots
X_{n-2}\}\equiv 0,\quad n\geq 2 .
\end{eqnarray}
Applying identity   (\ref{GenId}) to (\ref{FunmuII}),  we then
obtain identity (\ref{FunmII}). As in the case of (\ref{FunI}),
identity (\ref{GenId}) can be derive from (\ref{FunmII}). For this,
we sum  (\ref{FunmII}) over cyclic permutations. We have
\begin{eqnarray}
\label{FunmIII} \nonumber
&&\big((n-2)[X_1, X_2X_3\cdots X_n]-\{X_n,X_1X_2\cdots X_{n-1}\}+\{X_{n-1},X_nX_1\cdots X_{n-2}\}\big)+\\
&&\qquad +cycle(X_1,X_2,...,X_n)\equiv 0.
\end{eqnarray}
But
\begin{eqnarray}
\label{FunmIV} \nonumber \big(\{X_n,X_1X_2\cdots
X_{n-1}\}-\{X_{n-1},X_nX_1\cdots
X_{n-2}\}\big)+cycle(X_1,X_2,...,X_n)\equiv 0
\end{eqnarray}
and identity (\ref{GenId}) follows from (\ref{FunmIII}).  This
allows speaking about  identities (\ref{FunmII}) as fundamental for
any associative algebra.
\\

\section{Conclusions}

We have discussed identity (\ref{FunII}) for arbitrary associative
algebra and analog (\ref{FunIIs}) of this identity for an arbitrary
associative (super)algebra . We  proposed regarding these identities
as fundamental  because they are described in terms of single
commutators and anticommutators in contrast to the identities for
algebras usually discussed (see \cite{MR,LV}), which are  in fact
consequences of these identities.

We proved (Theorem 2 (Theorem 4)) that any algebra or superalgebra
endowed with  two bilinear operations (commutator and
anticommutator) satisfying identities (\ref{JIantiI}) and
(\ref{JIantiII}) or ((\ref{JIantis}) and (\ref{JIantiIs})) is an
associative algebra or superalgebra \footnote{Note that for algebras
this fact was proved in \cite{MR}.}. We can stress that there were
no assumptions in the proof concerning the finiteness or even the
existence of a basis for a given algebra or superalgebra.

 We  introduced identity (\ref{PBJI}) for any
nondegenerate even symplectic supermanifold and discussed an
application of this identity in the canonical quantization of
dynamical systems.

Finally, we proposed a generalization of the basic identities to the
case of an arbitrary number of elements
involved in these relations (see (\ref{GenId}) and (\ref{FunmII})).\\

\section*{Acknowledgments}
\noindent The authors thank V. V. Zharinov and  A. K. Pogrebkov  for
carefully reading the manuscript and also D. A. Leites and P. A.
Zusmanovich for the useful remarks and references concerning
identities in algebras.

 This work is supported in part by the Program for Supporting Leading Scientific Schools (grant 88.2014.2,
  P.M.L. and
 O.V.R.),
  the Ministry of Education and
Science of the Russian Federation (grant TSPU-122, P.M.L. and
 O.V.R), and the Russian Foundation for Basis Research (N 12-02-00121, P.M.L. and
 O.V.R., and grant N 14-01-00489, I.V.T.).

\end{document}